\documentclass[preprint,preprintnumbers,amsmath,amssymb,superscriptaddress]{revtex4}

\usepackage{graphicx}

\newcommand{\bsigma}{\mbox{\boldmath$\sigma$}}

\begin{document}

\title{Anderson localization casts clouds over adiabatic quantum optimization}
\author{\sc Boris Altshuler}\email{bla@phys.columbia.edu}
\affiliation{Columbia University}
\affiliation{NEC Laboratories America Inc.}
\author{\sc Hari Krovi}\email{hari.krovi@uconn.edu}
\affiliation{NEC Laboratories America Inc.}
\author{\sc Jeremie Roland}\email{jroland@nec-labs.com}
\affiliation{NEC Laboratories America Inc.}
\date{\today}

\begin{abstract}
Understanding NP-complete problems is a central topic in computer science. This is why adiabatic quantum optimization has attracted so much attention, as it provided a new approach to tackle NP-complete problems using a quantum computer. The efficiency of this approach is limited by small spectral gaps between the ground and excited states of the quantum computer's Hamiltonian. We show that the statistics of the gaps can be analyzed in a novel way, borrowed from the study of quantum disordered systems in statistical mechanics.
It turns out that due to a phenomenon similar to Anderson localization, exponentially small gaps appear close to the end of the adiabatic algorithm for large random instances of NP-complete problems. This implies that unfortunately, adiabatic quantum optimization fails: the system gets trapped in one of the numerous local minima.
\end{abstract}

\maketitle

\paragraph{NP-completeness.} One of the central concepts in computational complexity theory is that of NP-completeness~\cite{garey-johnson}. A computational problem belongs to the class NP if its solution can be verified in a time at most polynomial in the input size $N$, i.e., the verification requires not more than $cN^k$ computational steps, where $c$ and $k$ are independent of $N$. An NP-complete problem satisfies a second criterion: any other problem in the class NP can be reduced to it in polynomial time. Remarkably, such problems exist, many of them being of a great practical importance. The question of whether NP-complete problems are ``easy to solve'', or in other words whether they may be solved in polynomial time,  is one of the most fundamental open problems in computer science: this is the famous ``P $\stackrel{?}{=}$ NP'' question~\cite{arora-barak}. It is commonly believed however that it is not the case, i.e., that solving such a problem requires a computational time which is exponential in $N$.

\paragraph{Adiabatic quantum optimization.} The discovery of an efficient (polynomial time) quantum algorithm for the factorization of large numbers---a problem in NP but not believed to be NP-complete---is a milestone in quantum computing~\cite{Shor}, as no algorithm is known to solve this problem efficiently on a classical (non-quantum) computer. However, this success was not extended to NP-complete problems. That was why the proposal of Farhi {\it et al}.~\cite{farh00} to use adiabatic quantum optimization (AQO) to solve NP-complete problems has attracted much attention since initial numerical simulations suggested such a possibility~\cite{fggllp01}.

The basic idea of AQO is as follows: suppose that the solution of a computational problem $P$ can be encoded in the ground state (GS) of a Hamiltonian $\hat{H}_P$. To implement AQO one needs to construct a physical quantum system that is governed by a Hamiltonian $\hat{H}(s)=(1-s)\hat{H}_0 + s\hat{H}_P$ where $s$ is a tunable parameter, and $\hat{H}_0$ is a Hamiltonian with a known and easy-to-prepare ground state. The idea is to start with $s=0$, initialize the system in the ground state of $\hat{H}(0)=\hat{H}_0$ and increase $s$ with time as $s=t/T$. According to the Adiabatic Theorem~\cite{messiah}, slow enough variation of the parameter $s=s(t)$ keeps the system in the ground state of the Hamiltonian $\hat{H}(s(t))$ at any time $t$. Therefore, if $T$ is large enough, at $t=T$ the system would find itself in the ground state of $\hat{H}(1)=\hat{H}_P$ and the problem would be solved. This model has since been shown to be equivalent to the standard (circuit) model of quantum computing~\cite{avkll04}. Of course, as long as the computational time $T$ remains finite there is a non-zero probability that the system would undergo a Landau-Zener transition~\cite{messiah} and end up in an excited state. In order to maintain the excitation probability less than $\epsilon$, the adiabatic condition requires that $T\sim\frac{1}{\epsilon\Delta^2}$, where  $\Delta(s)=E_{ES}-E_{GS}$ is the energy gap between the ground state and first excited state (ES) of the Hamiltonian $\hat{H}(s)$. Therefore AQO is not efficient when $\Delta$ is small. More precisely, the adiabatic quantum approach to NP-complete problems can beat known classical algorithms (which require exponential time) provided that the minimal value of the gap scales as an inverse power of the problem size $N$.
Previously, it was shown that the gap can become exponentially small under specific conditions, such as a bad choice of initial Hamiltonian~\cite{znidaric05,Farhi_fail}, or for specifically designed hard instances~\cite{Reich,vdam01,vdam2}. In particular, it was recently argued that the presence of a first order phase transition could induce an exponentially small gap, and this effect was demonstrated for a particular instance of an NP-hard problem~\cite{AminChoi}, and later for {\it planted} instances of 3-SAT~\cite{mit}.
While these examples show that small gaps can occur for {\it specific} instances of NP-complete problems,
one could hope that this is not the typical behavior, i.e., for randomly generated instances the gap could be small only with very low probability. This hope followed from numerical simulations~\cite{fggllp01,hogg03,bolpr06} where the minimum gap seemed to decrease only polynomially for small instances, up to $N=124$ for the latest simulations~\cite{yks08}.
In this paper we show that this scaling does not persist for larger $N$. It turns out that as $N\rightarrow\infty$, the typical value of the minimal gap for random instances decays even faster than exponentially. As a result, the probability for AQO to yield a wrong solution in this limit tends to unity.

\begin{figure}
\begin{center}
\includegraphics[width=350pt]{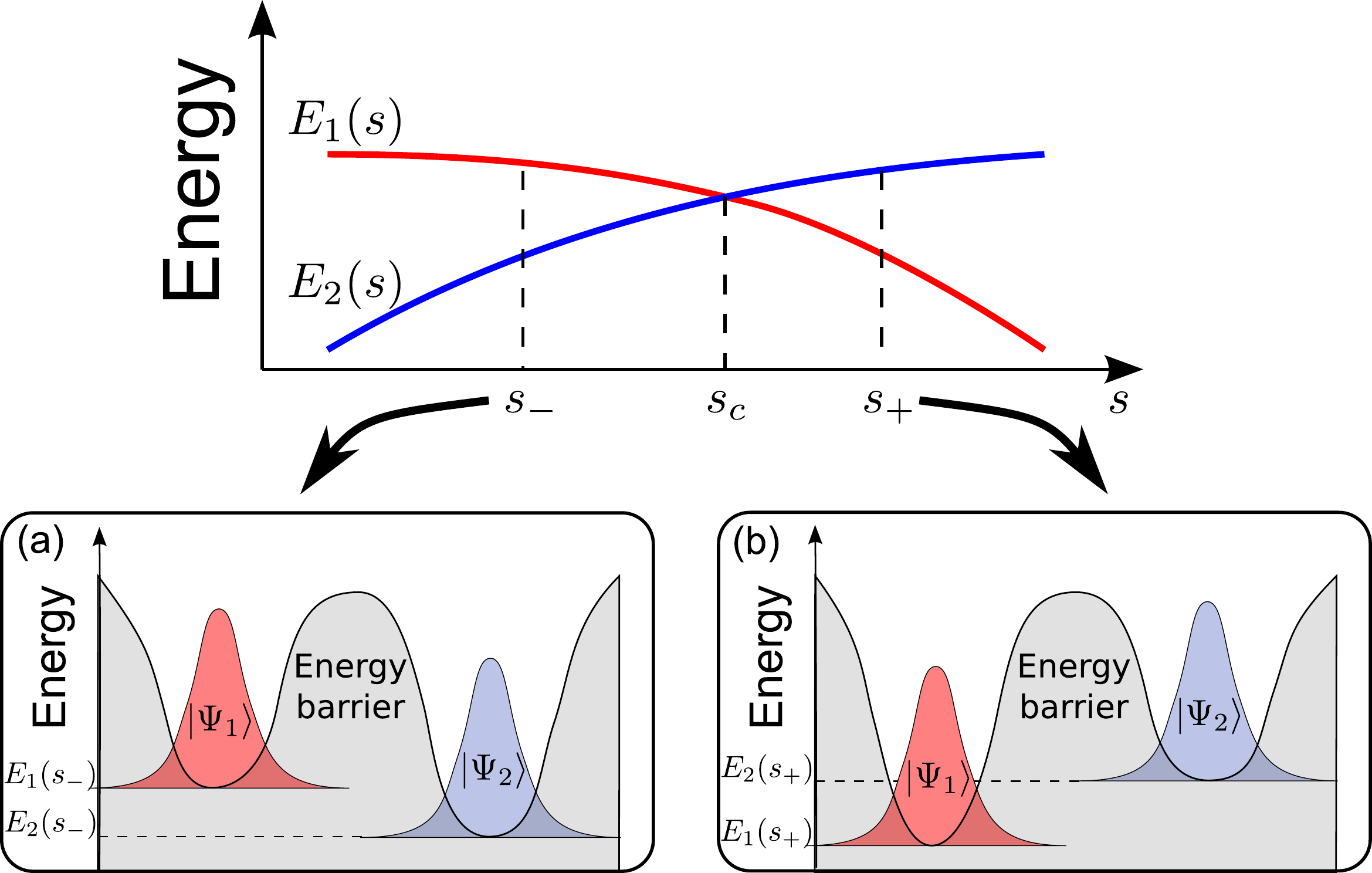}
\end{center}
\caption{Schematic representation of a level anti-crossing. The energies of two quantum states $|\Psi_1\rangle$ and $|\Psi_2\rangle$ localized in distant wells can be fine-tuned by applying a smooth additional potential. (a) Before the crossing, the ground state is $|\Psi_2\rangle$ with energy close to $E_2(s)$, i.e., for $s_-<s_c$, we have that $E_1(s_-)>E_2(s_-)$, so that $|GS(s_-)\rangle = |\Psi_2\rangle$. (b) After the crossing, the ground state becomes $|\Psi_1\rangle$ with energy close to $E_1(s)$, i.e. for $s_+>s_c$, we have that $E_1(s_+)<E_2(s_+)$, so that $|GS(s_+)\rangle = |\Psi_1\rangle$. The ground states before and after the crossing have nothing to do with each other. At a certain interval of $s$ close to $s_c$, the anti-crossing takes place and the ground state is a linear combination of $|\Psi_1\rangle$ and $|\Psi_2\rangle$.\label{fig:cartoon-anticrossing}}
\end{figure}

\paragraph{Anderson localization.} The appearance of exponentially small spectral gaps can be naturally attributed to the Anderson localization (AL) of the eigenfunctions of $\hat{H}(s)$ in the space of the solutions. Originally, AL implied that the wave function of a quantum particle in $d$-dimensional space ($d=1,2,3,\dots$) subject to a strong enough disorder potential turns out to be spatially localized in a small region and decays exponentially as a function of the distance from this region. Accordingly the probability for the particle to tunnel through a large disordered region is suppressed exponentially. To illustrate this, first note that the gap $\Delta$ can not vanish at any $s$ unless there is a special symmetry reason. This is the famous Wigner-von Neumann non-crossing rule~\cite{Wigner}: the curves that describe the $s$-dependence of two eigenenergies do not cross on the $(E,s)$-plane. This so-called level repulsion follows from the consideration of a reduced $2\times 2$ Hamiltonian that describes two anomalously close energy states and neglects the rest of the spectrum. Let $E_{1}$ and $E_{2}$ be the diagonal matrix elements of the Hamiltonian, and $V_{12} = V_{21}^*$ be its off-diagonal matrix elements.
We then find the energy gap to be
\begin{equation}\label{Energy_diff_2_level}
\Delta=E_{ES} - E_{GS} = \sqrt{(E_1-E_2)^2+|V_{12}|^2}.
\end{equation}
Now suppose that $E_1(s)$ and $E_2(s)$ become equal at $s=s_c$, as depicted in Fig.~\ref{fig:cartoon-anticrossing}.
We find that $\Delta>0$ even for $s=s_c$. This is known as a level anti-crossing. The minimal value of the energy gap is determined by the off-diagonal matrix element i.e., $\Delta_{\min}=|V_{12}|$ which is exponentially small under AL conditions. Accordingly the energy level repulsion between the localized states should be exponentially small in the spatial distance. Fig.~\ref{fig:cartoon-anticrossing} illustrates this situation schematically. At certain interval of $s$ close to $s_c$ the difference $E_1(s)-E_2(s)$ is smaller or of the order of the tunneling matrix element $V_{12}$. It is the interval where the anti-crossing takes place. Since $V_{12}$ depends exponentially on the distance between the wells, both the width of the anti-crossing interval and the minimum gap turn out to be exponentially small. The concept of AL was introduced more than 50 years ago in order to describe spin and charge transport in disordered solids~\cite{anderson58}. Since then AL was found to be relevant for a variety of physical situations. It also turned out to exist and make physical sense in a much broader class of spaces than $\mathcal{R}^d$. Below we demonstrate that a phenomenon analogous to AL on the vertices of the $N$-dimensional cube naturally appears in connection with AQO.

\paragraph{Exact Cover 3.} In order to explain the connection between the AQO approach to NP-complete problems and Anderson localization, we pick a particular NP-complete problem known as Exact Cover 3 (EC3), the same problem that was used for the early numerical simulations of AQO~\cite{fggllp01}. However, we believe that this analysis can be extended to any NP-complete problem. EC3 can be formalized in the following way. Consider $N$ bits $x_1, x_2, \dots , x_N$ which take values $0$ or $1$. An instance of EC3 consists of $M$ triplets of bit indices $(i_c,j_c,k_c)$ (the clauses), where each clause is said to be satisfied if and only if one of the corresponding bits is $1$ and the other two are $0$. A solution of a particular instance of EC3 is an assignment of the bits $\mathbf{x}=(x_1,x_2,\dots ,x_N)$ which satisfies all of the clauses. This problem can be assigned a cost function given by $f(\mathbf{x})=\sum_c(x_{i_c}+x_{j_c}+x_{k_c}-1)^2$: each solution has zero cost and all other assignments have a positive cost. We consider a standard distribution of random instances, where an instance is built by picking the $M$ clauses independently, each clause being obtained by picking 3 bit indices uniformly at random. The hardness of such random instances is characterized by the clauses-to-variables ratio $\alpha=M/N$. There are two characteristic values of $\alpha$: the clustering threshold $\alpha_{cl}$ and the satisfiability threshold $\alpha_s$~\cite{biroli}. For $\alpha<\alpha_{cl}$, the density of the solutions is high and essentially uniform, while for $\alpha>\alpha_{cl}$ the solutions become clustered in the solution space with different clusters remote from each other (the distance between two assignments is the so called Hamming distance which is defined as the number of bits in which they differ). As $\alpha$ increases from $\alpha_{cl}$ to $\alpha_s$, the clusters become smaller and the distance between them increases. For $\alpha>\alpha_s$, the probability that the problem is satisfiable vanishes in the limit $N,M\rightarrow\infty$. It has been shown~\cite{raymond07} that $\alpha_s\approx 0.6263$. We will be interested in instances with $\alpha$ close to $\alpha_s$, which only accept a few isolated solutions and are therefore hard to solve. More precisely, known classical algorithms can not solve such hard instances for a number of bits $N$ more than a few thousands, so that this is the regime where an efficient quantum algorithm would be particularly desirable.

\paragraph{Adiabatic quantum algorithm.} In order to define an adiabatic quantum algorithm for EC3, we need to choose $\hat{H}_P$ and $\hat{H}_0$. The problem Hamiltonian $\hat{H}_P$ for an EC3 instance can be obtained from the above cost function by first replacing $x_i$ by the Ising variables $\sigma_z^{(i)}=1-2x_i=\pm 1$ and then substituting $\sigma_z^{(i)}$ by the Pauli $Z$ operators $\hat{\sigma}_z^{(i)}$, thus replacing the bits by qubits. The problem Hamiltonian becomes 
\begin{equation}\label{Eq:H_P}
\hat{H}_P= M\hat{I} - \frac{1}{2}\sum_{i=1}^N B_i\hat{\sigma}_z^{(i)} + \frac{1}{4}\sum_{i,j=1}^N J_{ij} \hat{\sigma}_z^{(i)}\hat{\sigma}_z^{(j)},
\end{equation}
where $B_i$ is the number of clauses that involve the bit $i$, $J_{ij}$ is the number of clauses where the bits $i$ and $j$ participate together, and $\hat{I}$ is the identity operator. For $\hat{H}_0$, we make the conventional choice $\hat{H}_0=-\sum_i \hat{\sigma}_x^{(i)}$, which corresponds to spins in the magnetic field directed along $x$-axis (Pauli $X$ operators). 
For us it will also be convenient to modify the Hamiltonian $\hat{H}(s)$ as $\hat{H}_{QC}(\lambda)=\hat{H}_P+\lambda\hat{H}_0$. The parameter $\lambda=\frac{1-s}{s}$ changes adiabatically from $\lambda=+\infty$ at the beginning $t=0$ to $\lambda=0$ at $t=T$.

\paragraph{Connection to Anderson Localization.} We can now see the relevance of AL to the quantum system described by $\hat{H}_{QC}$. Note that this Hamiltonian also describes a single quantum particle that is moving between the vertices of an $N$-dimensional hypercube. Indeed, each vector $\bsigma=(\sigma_z^{(1)},\sigma_z^{(2)},\dots,\sigma_z^{(N)})$, where $\sigma_z^{(i)}=\pm 1$, determines a vertex of the hypercube, which is body-centered at the origin of the $N$-dimensional space. Let $|\bsigma\rangle$ denote the quantum state of a particle localized at a site $\bsigma$. The full set of these states forms a basis, in which the first term of the Hamiltonian is diagonal, while the second one describes a hopping of this fictitious particle between the nearest neighbors (n.n)
\begin{equation}\label{Eq:AndersonModel}
\hat{H}_{QC}(\lambda)=\underbrace{\sum_{\bsigma}E_P(\bsigma) |\bsigma\rangle\langle \bsigma |}_{\textrm{disorder}} + \lambda \sum_{\bsigma,\bsigma^\prime\ n.n} |\bsigma\rangle\langle \bsigma^\prime|.
\end{equation}
Each on-site energy $E_P(\bsigma)$ is nothing but the cost function $f(\mathbf{x})$ of the corresponding assignment $\bsigma$. For random instances, the on-site energies are obviously also random, introducing disorder in the Hamiltonian. Hence, Eq.~(\ref{Eq:AndersonModel}) describes the well known Anderson model, which was used to demonstrate the phenomenon of localization~\cite{anderson58}. The only difference from more familiar situations is that lattices in $d$-dimensional space, which have $L^d$ sites where $L\gg 1$ is the system size, are substituted by the $N$-dimensional hypercube with $2^N$ sites, where $N\gg 1$. 

\begin{figure}
\begin{center}
\includegraphics[width=225pt]{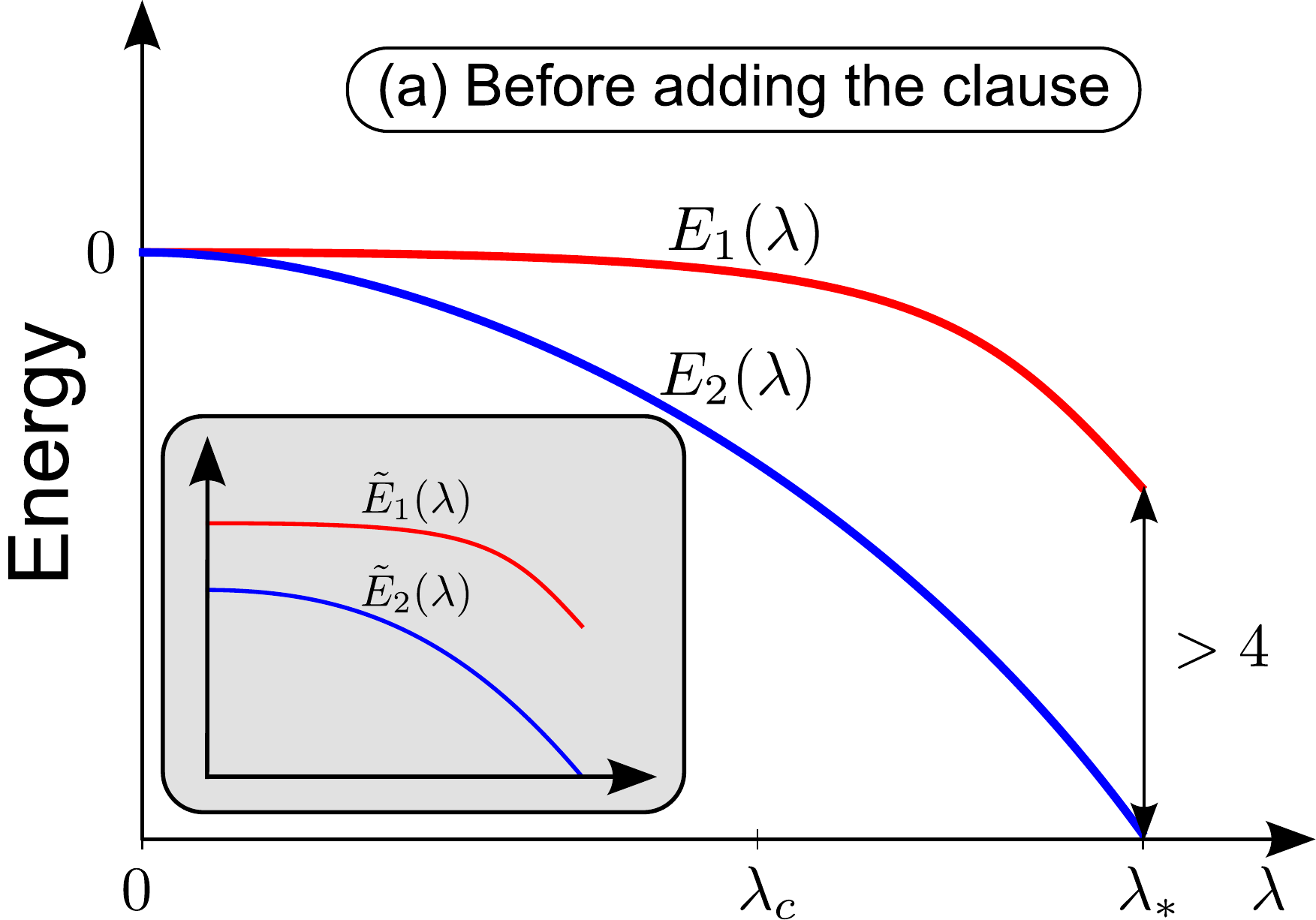}
\includegraphics[width=225pt]{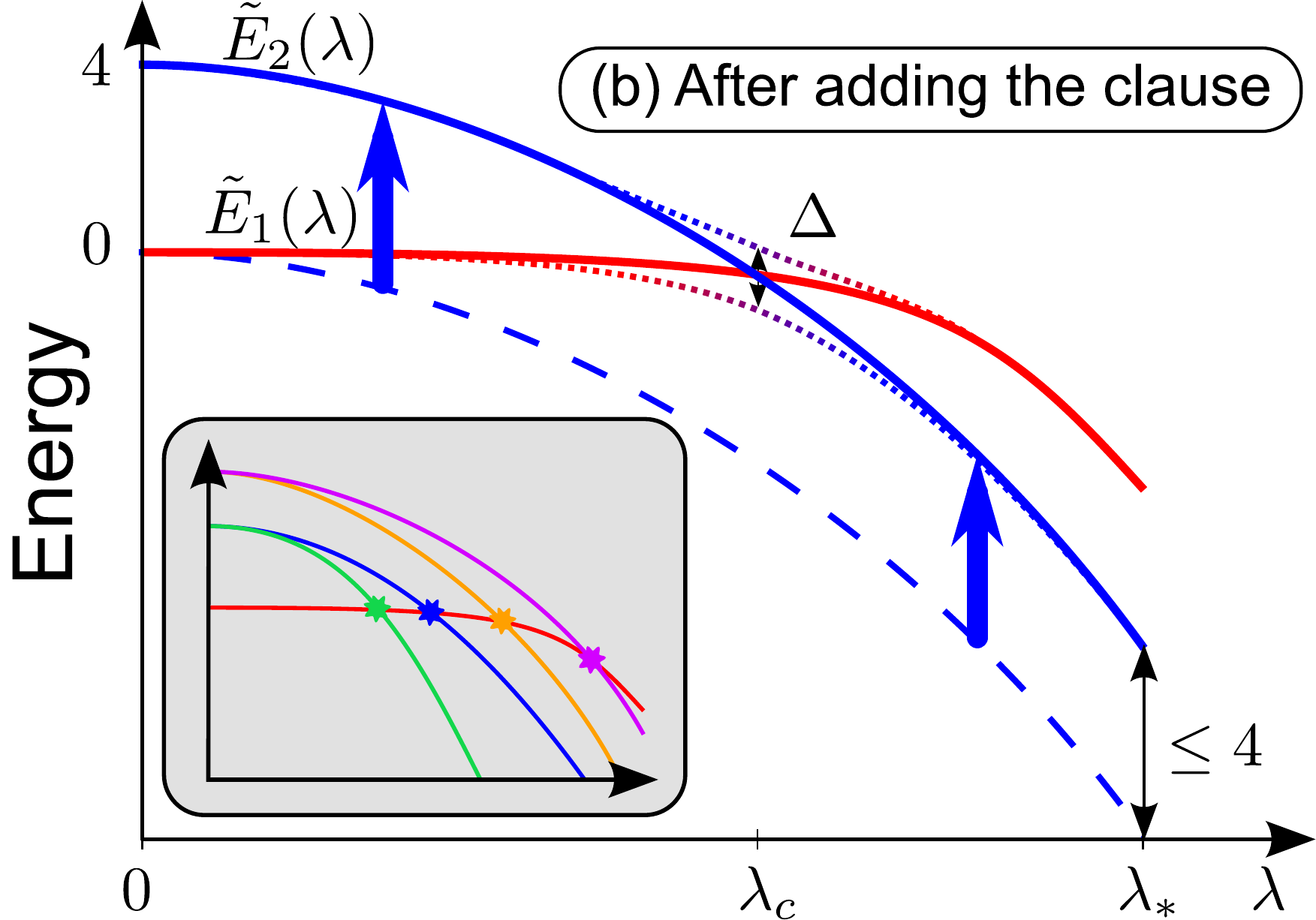}
\end{center}
\caption{Schematic representation of the creation of a level anti-crossing. (a) Before adding the clause, we have two assignments which are both in the ground state at $\lambda=0$ but due to the no-crossing rule, at $\lambda>0$ we have $E_1(\lambda_*)-E_2(\lambda_*)>4$. (b) By adding a clause satisfied by solution 1 but not solution 2, we create a level anti-crossing since $\tilde{E}_1(0)<\tilde{E}_2(0)$ but $\tilde{E}_1(\lambda_*)>\tilde{E}_2(\lambda_*)$. Insets: (a) If the clause is violated by the wrong solution, then no anti-crossing appears between these two levels. (b) However, other low energy levels can create other anti-crossings, leading to multiple small gaps.\label{fig:level-crossing}}
\end{figure}

\paragraph{Anti-crossings in AQO.} Now we are ready to discuss the fundamental difficulties which AQO faces. We will show that (i) the anti-crossings of the ground state with the first excited state happen with high probability and (ii) that the anti-crossing gaps in the limit $N\rightarrow\infty$ are even less than exponentially small. Let us start with the first statement. An EC3 instance with $\alpha<\alpha_s$ typically has several solutions $\bsigma$ with $E_P(\bsigma)=0$. If $\alpha$ is close to $\alpha_s$ there are few solutions at a distance of order $N$ of each other. The presence of multiple solutions imply that the ground state of $\hat{H}_{QC}(\lambda=0)=\hat{H}_P$ is degenerate, but this does not contradict the non-crossing rule: $\hat{H}_P$ commutes with each of the operators $\hat{\sigma}_z^{(i)}$, so it satisfies a special symmetry which is broken for $\lambda>0$. Consider now a particular instance with $M-1$ clauses accepting two solutions $\bsigma_1$ and $\bsigma_2$ that are separated by $n\sim N$ spin flips. When $\lambda$ adiabatically changes from zero to a small but finite value the solutions evolve into eigenstates of the Hamiltonian, $|\Psi_1,\lambda\rangle$ and  $|\Psi_2,\lambda\rangle$ with the energies $E_1(\lambda)$ and $E_2(\lambda)$. According to the non-crossing rule, a degeneracy of these two states at a finite $\lambda$ is improbable, i.e., the $\hat{H}_0$ term in $\hat{H}_{QC}$ splits the ground state degeneracy. This situation is sketched in Fig.~\ref{fig:level-crossing}(a).
Suppose that $E_2(\lambda)<E_1(\lambda)$, i.e. $|\Psi_2,\lambda\rangle$ is the unique ground state of the Hamiltonian $\hat{H}_{QC}(\lambda)$. If we now add one more clause to the existing $M-1$ ones, i.e. we add a term $(x_{i_M}+x_{j_M}+x_{k_M}-1)^2$
to the cost function leading to Hamiltonian $\hat{H}_P$, both $|\bsigma_1\rangle$ and $|\bsigma_2\rangle$ remain eigenstates, but their eigenenergy can increase by either 1 or 4. With a non-zero probability the last clause is satisfied by $\bsigma_1$ but not by $\bsigma_2$, i.e., $\tilde{E}_P(\bsigma_1)=0$ while $\tilde{E}_P(\bsigma_2)>0$, where $\tilde{E}_P(\bsigma)$ is the cost function of the new instance. Accordingly $|\bsigma_1\rangle$ rather than $|\bsigma_2\rangle$ is the new ground state at $\lambda=0$. At the same time $|\Psi_2,\lambda\rangle$ can still remain the ground state at large enough $\lambda$ if $\tilde{E}_1(\lambda)>\tilde{E}_2(\lambda)$, as shown on Fig.~\ref{fig:level-crossing}(b). Such a situation corresponds to the anti-crossing of $|\Psi_1,\lambda\rangle$ and $|\Psi_2,\lambda\rangle$ at certain $\lambda$, as previously described in Fig.~\ref{fig:cartoon-anticrossing}. Note that the addition of a clause
to the cost function increases any eigenenergy of $\hat{H}_{QC}(\lambda)$ by less than 4. To satisfy the condition $\tilde{E}_1(\lambda)>\tilde{E}_2(\lambda)$, it is thus sufficient to achieve a large enough splitting between the eigenvalues of the instance with $M-1$ clauses: $E_1(\lambda)-E_2(\lambda)>4$. It turns out that if $N\gg 1$, this happens when $\lambda$ is small and one can use perturbation theory in $\lambda$.

\begin{figure}
\begin{center}
\includegraphics[width=250pt]{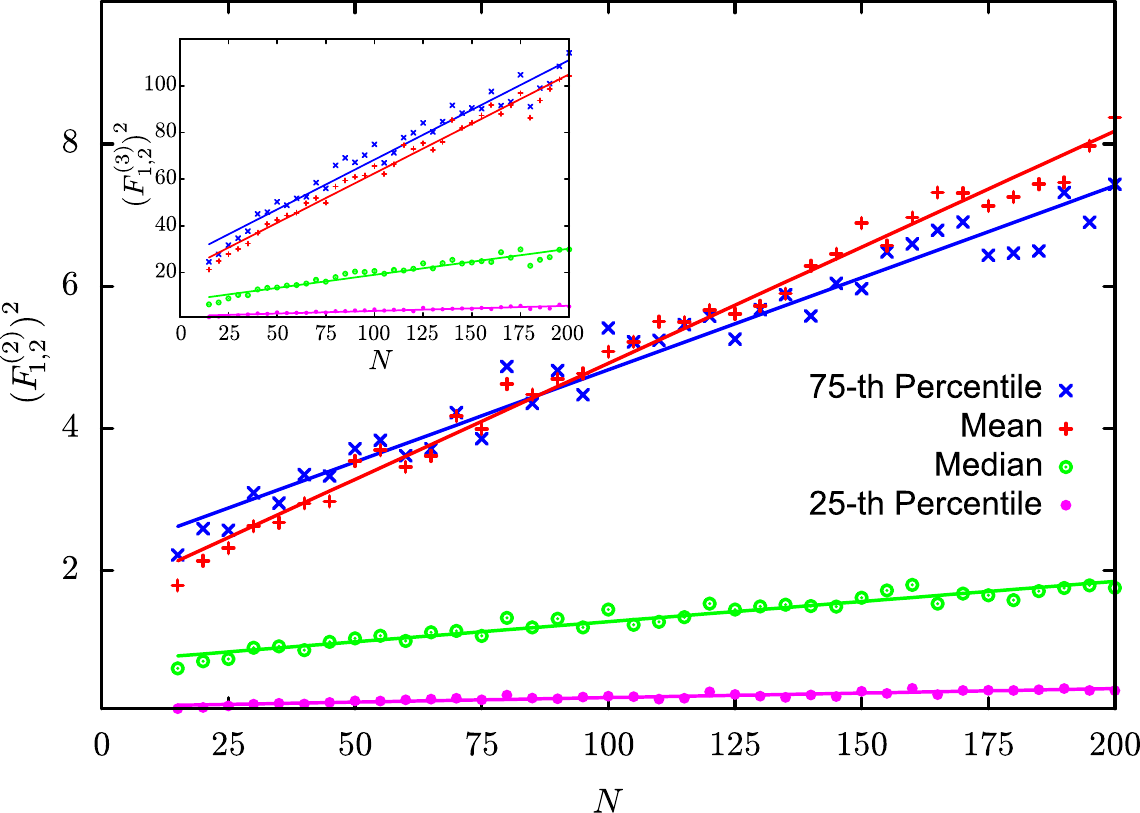}\hspace*{0.5cm}
\end{center}
 \caption{Statistics of the square of the difference in energies of two solutions up to fourth order i.e. $(F^{(2)}_{1,2})^2$. Linear fits confirm that the square of the energy difference scales as $O(N)$. Inset: Statistics of the sixth order correction of the splitting $(F^{(3)}_{1,2})^2$. Each data point is obtained from 2500 random instances of EC3 with $\alpha\approx 0.62$. Linear fits for the mean yield $f^{(2)}\approx 0.18$ and $f^{(3)}\approx 0.65$.\label{fig:numerical-final}}
\end{figure}

\begin{figure}
\begin{center}
\includegraphics[width=250pt]{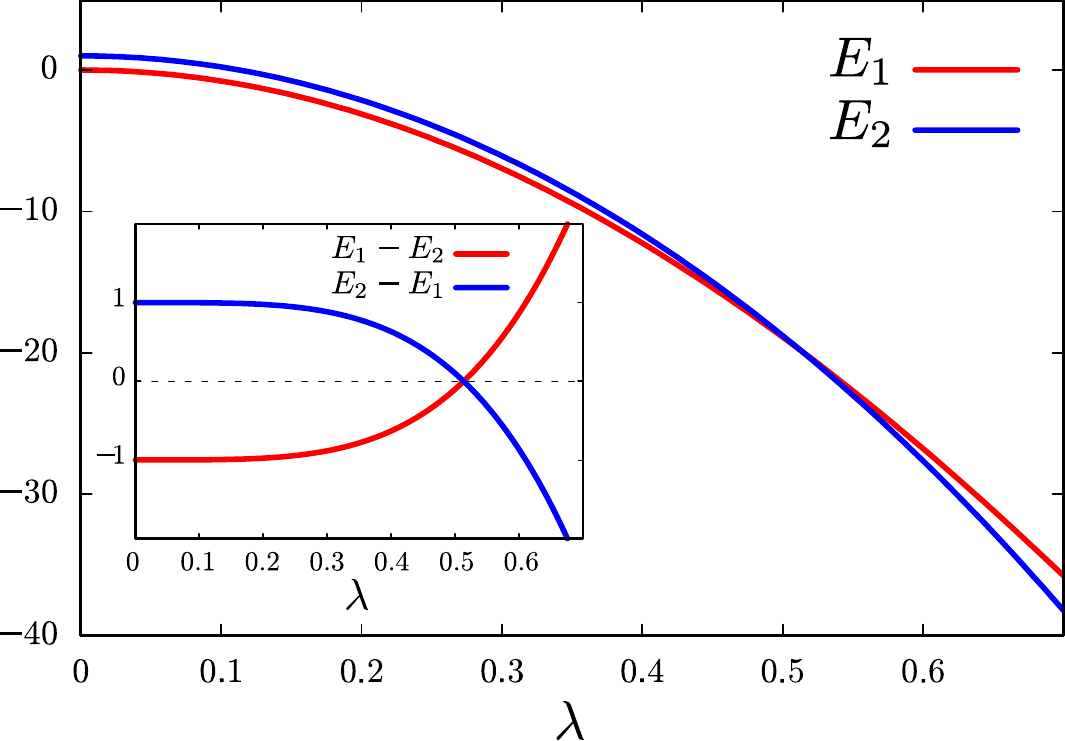}\hspace*{0.5cm}
\end{center}
 \caption{Simulation of a level anti-crossing for a random instance with $N=200$ bits and $\alpha\approx 0.62$, obtained by fourth order perturbation theory. The figure shows the energies of two assignments after adding a clause to the final Hamiltonian. The added clause is satisfied by assignment 1 but not by assignment 2. The figure shows a level crossing similar to the cartoon in Fig.~\ref{fig:level-crossing}. Inset: To make the crossing more apparent, we plotted the energy differences $E_1-E_2$ and $E_2-E_1$. The crossing occurs at $\lambda\approx 0.51$, and the corresponding assignments are at distance $n=60$ from each other.\label{fig:simulated-crossing-final}}
\end{figure}

\paragraph{Perturbation theory.} To demonstrate this, consider the eigenstate which in the limit $\lambda\rightarrow 0$ evolves to $|\bsigma\rangle$. At small $\lambda$ its energy can be expanded in a series
\begin{equation}\label{Eq:Energy_exp}
E(\lambda,\bsigma)=E_P(\bsigma) + \sum_{m=1}^\infty \lambda^{2m}F^{(m)}(\bsigma) .
\end{equation}
We can show that each term in this sum scales linearly in $N$. For the energy $E_P(\bsigma)$ of an arbitrary assignment, we immediately have that $0\leq E_P<M=\alpha N$. As for the coefficients $F^{(m)}(\bsigma)$, the cluster expansion~\cite{mayer} of the Hamiltonian $\hat{H}_{QC}$ implies that they may be expressed as a sum of $\sim N$ statistically independent terms, each being of order 1. The key element to prove this is that since $M/N =\alpha$ is constant, with high probability each bit participates in a finite number of clauses as $N\rightarrow\infty$. As a result, all the coefficients $B_i$ and $J_{ij}$ in Eq.~(\ref{Eq:H_P}) are also finite: $B_i=\sum_jJ_{ij}=O(1)$. In particular, when $\bsigma$ is a solution we obtain $F^{(1)}(\bsigma)=\sum_iB_i^{-1}$, which is therefore of order $N$.
This statement is valid for $F^{(m)}(\bsigma)$ with arbitrary finite $m>1$: all of them can be presented as a finite sum of $O(N)$ random terms, each one being of order unity.
Let us now consider the perturbative expansion for the energy splitting between two solutions. Similarly to Eq.~(\ref{Eq:Energy_exp}), we obtain
\begin{equation}
E_1(\lambda) - E_2(\lambda) = \sum_{m=2}^\infty \lambda^{2m}F^{(m)}_{1,2} ,
\end{equation}
where $F^{(m)}_{1,2}=F^{(m)}(\bsigma_1)-F^{(m)}(\bsigma_2)$ is a sum of $O(N)$ terms of order 1. Each of the terms is random with a zero mean and hence the sums $F^{(m)}_{1,2}$ averages to zero if $N$ is large. Therefore, it is $(F^{(m)}_{1,2})^2$ rather than $F^{(m)}_{1,2}$ which is proportional to $N$. We thus arrive to the conclusion that
\begin{equation}\label{Eq:Energy_diff}
|E_1(\lambda) - E_2(\lambda)| = \sqrt{N}\sum_m\lambda^{2m}f^{(m)} ,
\end{equation}
where the coefficients $f^{(m)}=O(1)$
can be evaluated by the cluster expansion~\cite{mayer}. We have seen that $F^{(1)}(\bsigma)=\sum_iB_i^{-1}$ for any solution $\bsigma$, so that $F^{(1)}_{1,2}=0$. However terms with $m>1$ do not vanish, making the splitting finite. On Fig.~\ref{fig:numerical-final}, we show the results of the statistical analysis of the numerical calculations of the coefficients $(F^{(2)}_{1,2})^2$ and $(F^{(3)}_{1,2})^2$, with linear fits confirming their scaling $O(N)$.
For small $\lambda$, we can restrict ourselves to the leading term ($m=2$) in Eq.~(\ref{Eq:Energy_diff}). Accordingly in the $N\rightarrow\infty$ limit, the splitting $|E_1(\lambda) - E_2(\lambda)|$ exceeds $4$ as long as $\lambda>\lambda_\ast$, with
\begin{equation}\label{Eq:lambda_star}
\lambda_\ast = \sqrt{2}\ (f^{(2)})^{-1/4}\ N^{-1/8},
\end{equation}
and $\lambda_\ast\ll 1$ so that we can neglect higher orders, $\lambda_*\ll 1$ (the validity of this approximation will be discussed in the next paragraph). From Eq.~(\ref{Eq:lambda_star}), it follows that the anti-crossing probability for the instance with $M$ clauses is finite provided that $\lambda \geq\lambda_\ast \sim N^{-1/8}$.
How big is the gap $\Delta$ of such an anti-crossing? As explained above, we can evaluate the gap by considering the matrix element $V_{12}$ between the states $|\Psi_1,\lambda\rangle$ and $|\Psi_2,\lambda\rangle$ corresponding to the two assignments, at the value $\lambda$ where the anti-crossing occurs. Note that if the two assignments $\bsigma_1$ and $\bsigma_2$ satisfying the $(M-1)$ clauses are separated by a distance (number of flips) $n$, this matrix element only appears at the $n$-th order of the perturbation theory, i.e. it is proportional to $\lambda^{n}$:
\begin{equation}\label{Eq:Matrix_element}
V_{12} = \lambda^n\sum_{tr}\left(\Pi_{k=1}^n E_P(\bsigma_{tr}^{(k)})\right)^{-1} + O(\lambda^{n+1})
\end{equation}
where the sum is over all "trajectories" $tr$ - all possible orders of the $n$ spin flips needed to transform $\bsigma_1$ into $\bsigma_2$, $\bsigma_{tr}^{(k)}$ is the assignment along a particular trajectory that appears after $k$ flips and $E_P(\bsigma_{tr}^{(k)})$ is the cost function of this assignment.
Therefore we can estimate the matrix element and thus the anti-crossing gap as $V_{12}<w(n)\lambda^{n}$. The prefactor $w(n)$ reflects the fact that many ($\sim n!$) trajectories contribute to the sum in Eq.~(\ref{Eq:Matrix_element}). For a typical trajectory $E_P(\bsigma_{tr}^{(k)})=O(k)$ for $k<n/2$ and $E_P(\bsigma_{tr}^{(k)})=O(n/2-k)$ for $k>n/2$. As a result the product of $E_P(\bsigma_{tr}^{(k)})$ in Eq.~(\ref{Eq:Matrix_element}) is also $\sim n!$.
The factorials thus cancel each other and $w(n)$ can not increase faster than $A^n$ with some constant $A\sim 1$. Therefore, $V_{12}<(A\lambda)^n$. Combining this with Eq.~(\ref{Eq:lambda_star}), we see that an anti-crossing at $\lambda$ close to $\lambda_*$ yields the minimum gap
as small as $\Delta_{\min}\sim\exp [-(n/8)\ln(N/N_0)]$, where $N_0=16A^{8}(f^{(2)})^{-2}=O(1)$.
We can estimate the distance $n$ between the assignments as $v(\alpha)N$, where $v(\alpha)\approx (4/9)(1-\exp(-3\alpha))$, and obtain the final form of the minimal gap estimation 
\begin{equation}\label{Eq:Minimum_gap}
\Delta_{\min}\sim\exp [(-v(\alpha)N/8)\ln (N/N_0)]. 
\end{equation}
One can see that as $N\rightarrow\infty$, the gap indeed decreases even faster than an exponential - statement (ii). This implies that the adiabatic computation time exceeds $\exp(N)$. In Fig.~\ref{fig:simulated-crossing-final}, we have plotted an anti-crossing  for a particular instance with $N=200$ generated during our numerical simulations. The figure shows two energy levels (estimated by fourth order perturbation theory) corresponding to assignments separated by 60 bit flips, and crossing at $\lambda\approx 0.51$.

\paragraph{Applicability of the perturbation theory.} Our main result - the estimation of the minimal gap (Eq.~(\ref{Eq:Minimum_gap})), is based on the perturbative expansion for the energies (Eq.~(\ref{Eq:Energy_exp})) and the matrix element $V_{12}$ (Eq.~(\ref{Eq:Matrix_element})). Is the perturbation theory in $\lambda$ always applicable? At first sight Eq.~(\ref{Eq:Matrix_element}) becomes meaningless if $E_P=0$ for any of the intermediate assignments $\bsigma_{tr}^{(k)}$. In this case there is an avoided crossing between the states corresponding to the assignments $\bsigma_1$ and $\bsigma_{tr}^{(k)}$ (such as in Eq.~(\ref{Energy_diff_2_level})) and formally perturbation theory fails in the vicinity of this anti-crossing point. This apparent difficulty can be overcome by considering only a finite time $T$ for the evolution. This is equivalent to adding imaginary parts $i\eta\approx i/T$ to the energies. For the AQO algorithm, it is the computation time $T$ that determines $\eta$. Since we are considering the $N\to\infty$ limit, we have that $T\to\infty$ and thus $\eta\to 0$. This is the limit that was shown to be relevant for the localization problem~\cite{anderson58,anderson-nobel}.
The celebrated discovery of Anderson was that if the limit $\eta\to 0$ is taken after the volume (here $N$) tends to infinity, and $\lambda$ is small enough i.e., $\lambda<\lambda_{cr}$, the spectrum of the Hamiltonian described in Eq.~(\ref{Eq:AndersonModel}) remains discrete (all states are localized) and thus the second term in Eq.~(\ref{Eq:AndersonModel}) (the kinetic energy term) can be treated perturbatively. As soon as $\lambda >\lambda_{cr}$, there appears a strip of extended states in the middle of the energy band which widens as $\lambda$ increases further. States within this strip are not perturbative because the number of the trajectories connecting two points in a $d$-dimensional space (for finite $d$) increases exponentially with distance. The large number of terms in the expansions like Eq.~(\ref{Eq:Matrix_element}) overwhelms the smallness of $\lambda^n$ and the perturbation series thus diverges for $\lambda>\lambda_{cr}$. For a $d$-dimensional space, the critical value $\lambda_{cr}$ is believed to be (in our units) of the order of $\lambda_{cr}\sim 1/\log d$~\cite{efetov,abou-chacra}. 
We have seen that the AQO algorithm for problems like EC3 can be mapped to the Anderson model on an $N$-dimensional hypercube. Then, the number of trajectories increases with the length $n$ as $n!\sim n^n e^{-n}$, i.e. even faster than an exponential.  However, as we already mentioned, the $n^n$ factor cancels with the same factor in the products of the energy in the denominators of Eq.~(\ref{Eq:Matrix_element}).  Accordingly, $\lambda_{cr}$ can still be estimated as $\lambda_{cr}\sim 1/\log N$, which, together with Eq.~(\ref{Eq:lambda_star}), implies that anti-crossings appear for  $\lambda_*\ll\lambda_{cr}$ when $N\gg 1$. Moreover, at $\lambda<\lambda_{cr}$ all of the states are supposed to be localized. The AQO algorithm involves only low energy states, which remain localized much longer than the middle-band states with the energies $\sim N$. Therefore, it is quite likely that the exponentially small gaps appear even at $\lambda\sim 1$.

\paragraph{Conclusions.} We finish our discussion with the following observation. We monitored two assignments that satisfied $M-1$ clauses and added an extra clause to create a small gap at finite $\lambda$. Of course, for randomly selected clauses this happens only with a finite probability and the situation sketched in the inset in Fig.~\ref{fig:level-crossing}(a) is also possible. One could thus hope~\cite{mit} that the AQO algorithm can find the solution with a sizable probability. Unfortunately, this is not the case. Indeed, let us adopt the most conservative limitation on the perturbative approach $\lambda_{cr}\sim 1/\log N$ and consider the spectrum at $\lambda_*\ll\lambda_{cr}\sim 1/\log N$. According to Eq.~(\ref{Eq:Energy_diff}) all states in the energy interval $[0,\epsilon]$ with $\epsilon\sim\sqrt{N}\lambda^4\gg 1$ have similar chances to evolve into the ground state at $\lambda=0$. This means that typically the ground state undergoes $\nu(\epsilon)$ anti-crossings (participates in $\nu(\epsilon)$ anti-crossing gaps) as the parameter evolves from $0$ to $\lambda$ (see the inset of Fig.~\ref{fig:level-crossing}(b)). Here $\nu(\epsilon)$ is the number of states, whose energies at the given $\lambda$ differ from the ground state energy by less than $\epsilon$. Taking into account that $\nu(\epsilon)$ increases with $\epsilon$ exponentially and that the probability to completely avoid anti-crossings (the probability to have a gap of size $\epsilon$ separating the ground state from the rest of the spectrum) is exponentially small in $\nu(\epsilon)$ we conclude that this probability is indeed negligible. Therefore, these findings suggest that there is no chance of obtaining the solution of the problem in polynomial time using the AQO algorithm for random instances of the Exact Cover 3 problem. We also believe that the methods described in this article can be applied to other similar NP-complete problems, such as 3-SAT.

\begin{acknowledgments}
 We thank A. Childs, E. Farhi, J. Goldstone, S. Gutmann, M. R\"otteler and A. P. Young for interesting discussions. We thank the \emph{High Availability Grid Storage} department of NEC Laboratories America for giving us access to their cloud to run our numerical simulations. This research was supported in part by US DOE contract No. AC0206CH11357.
\end{acknowledgments}

\bibliography{adiabatic_short_paper}

\end{document}